\documentclass[pra,twocolumn,showpacs,preprintnumbers,amsmath,amssymb]{revtex4}
\usepackage{graphicx}% Include figure files
\usepackage{dcolumn}% Align table columns on decimal point
\usepackage{bm}% bold math
\usepackage{subfigure}%
\usepackage{multirow}
\usepackage{booktabs}
\usepackage{amsthm,amsmath,amssymb,fancyhdr,color,graphicx}
\usepackage[verbose]{wrapfig}
\usepackage{picinpar,booktabs}

\begin{document}

%\preprint{APS/123-QED}

\title{ Non-Gaussian postselection and virtual photon subtraction in continuous-variable quantum key distribution }% Force line breaks with \\

\author{Zhengyu Li$^{1,2}$}
\author{Yichen Zhang$^3$}
\author{Xiangyu Wang$^3$}
\author{Bingjie Xu$^2$}
\author{Xiang Peng$^{1 \dag }$}
\author{Hong Guo$^{1}$}
\thanks{Corresponding author: hongguo@pku.edu.cn.
\\$^{\dag}$Corresponding author: xiangpeng@pku.edu.cn.}

\affiliation{$^1$State Key Laboratory of Advanced Optical Communication Systems and Networks, Center for Computational Science and Engineering and Center for Quantum Information Technology, School of Electronics Engineering and Computer Science, Peking University, Beijing 100871, China}
%\affiliation{$^2$, Peking University, Beijing 100871, China}
\affiliation{$^2$Science and Technology on Security Communication Laboratory, Institute of Southwestern Communication, Chengdu 610041, China}
\affiliation{$^3$State Key Laboratory of Information Photonics and Optical Communications, Beijing University of Posts and Telecommunications, Beijing 100876, China}

\date{\today}% It is always \today, today,
             % but any date may be explicitly specified

\begin{abstract}

Photon subtraction can enhance the performance of continuous-variable quantum key distribution (CV QKD). However, the enhancement effect will be reduced by the imperfections of practical devices, especially the limited efficiency of a single-photon detector. In this paper, we propose a \emph{non-Gaussian} postselection method to emulate the photon substraction used in coherent-state CV QKD protocols.
The \emph{virtual} photon subtraction not only can avoid the complexity and imperfections of a practical photon-subtraction operation, which extends the secure transmission distance as the ideal case does, but also can be adjusted flexibly according to the channel parameters to optimize the performance.
Furthermore, our preliminary tests on the information reconciliation suggest that in the low signal-to-noise ratio regime, the performance of reconciliating the postselected non-Gaussian data is better than that of the Gaussian data, which implies the feasibility of implementing this method practically.

\end{abstract}

\pacs{03.67.Dd, 03.67.Hk}% PACS, the Physics and Astronomy
\maketitle

\section{Introduction}
Quantum key distribution (QKD) \cite{RevModPhys.74.145.2002, RevModPhys.81.1301.2009} is the most applicable technology of quantum information, which can allow two users (Alice and Bob) to establish secure keys remotely through an insecure quantum channel controlled by an eavesdropper (Eve). QKD has two main branches, i.e., discrete-variable (DV) QKD
and continuous-variable (CV) QKD \cite{Rev.Mod.Phys.77.513.2005,Phys.Rep.448.1.2007,Rev.Mod.Phys.84.621.2012}, in which the information is carried by the quadratures ($\hat x$ and $\hat p$) of the light field. CV QKD protocols using Gaussian modulated coherent states \cite{Phys.Rev.Lett.88.057902.2002,Nature.421.238.2003,Phys.Rev.Lett.93.170504.2004} not only have been proved to be unconditionally secure in theory \cite{Phys.Rev.Lett.97.190502.2006,Phys.Rev.Lett.97.190503.2006,Phys.Rev.Lett.102.110504.2009,Phys.Rev.Lett.110.030502.2013, Phys.Rev.Lett.114.070501.2015}, but also have the advantage of being compatible with standard telecommunication technology, which leads to an expectation of better application.
However, limited by the practical experimental techniques and the non-perfect reconciliation efficiency, the transmission distances of early CV QKD setups were not sufficiently long for network applications \cite{Phys.Rev.A.76.042305.2007,New.J.Phys.11.045023.2009,Opt.Express.20.14030.2012}. Thus, research on extending the transmission distance has attracted much attention in the past few years.

Recent major progress in CV QKD in experiment \cite{Nat.Photonics.7.378.2013} was achieved with 80~km transmission distance using coherent states by taking advantage of the multidimensional reconciliation protocol \cite{Phys.Rev.A.77.042325.2008, Phys.Rev.A.84.062317.2011} in the low signal-to-noise ratio (SNR) regime and the optimization of other experimental aspects. Besides the improvement of experimental techniques, quantum operations were also proposed to improve the performance of CV QKD, such as the noiseless linear amplification (NLA) operation \cite{T.C.Ralph_NLA_QCMC_2009,Phys.Rev.A.86.012327.2012,Phys.Rev.A.87.062311.2013}. It can increase the transmission distance roughly by the equivalent of~$20{\log _{10}}g$~dB losses, where $g$ is the gain of the NLA.
 %Several linear optics implementations of NLA have been practically realized for the amplification of coherent states \cite{Nat.Photonics.4.316.2010, Phys.Rev.Lett.104.123603.2010, Phys.Rev.A.83.063801.2011, Nat.Photonics.5.52.2011}.
 Furthermore, to avoid the difficulty of sophisticated physical NLA operations \cite{Nat.Photonics.4.316.2010, Phys.Rev.Lett.104.123603.2010, Phys.Rev.A.83.063801.2011, Nat.Photonics.5.52.2011}, non-deterministic \emph{virtual} NLA via Gaussian postselection has been proposed \cite{Phys.Rev.A.86.060302.2012,Phys.Rev.A.87.020303.2013} and experimentally implemented \cite{Nat.Photonics.8.333.2014}.
%The main reason for the performance improvement of employing the NLA is that, the NLA can distill the entanglement of TMSV.

%On the other hand, photon subtraction, a non-Gaussian operation, can also enhance the entanglement of a TMSV source\cite{Phys.Rev.A.61.032302.2000, Phys.Rev.A.71.043805.2005, Phys.Rev.A.73.042310.2006, Phys.Rev.A.86.012328.2012}.

Alternatively, a photon-subtraction operation \cite{Phys.Rev.A.61.032302.2000, Phys.Rev.A.71.043805.2005, Phys.Rev.A.73.042310.2006, Phys.Rev.A.86.012328.2012} is shown to be able to significantly improve the transmission distance of CV QKD protocols using two-mode squeezed vacuum (TMSV) states \cite{Phys.Rev.A.87.012317.2013}. By exploiting the equivalence between the entanglement-based (EB) scheme and the prepare-and-measure (PM) scheme, it can also be employed in protocols using coherent states. However, the improvement will be reduced by the imperfections of devices in a practical photon-subtraction operation, especially the single-photon detector (SPD), which makes this method unfeasible (see Appendix \ref{imperfect}).
%Various scenarios were proposed, including applying PS on one mode or on both two modes of the TMSV \cite{Phys.Rev.A.73.042310.2006}, directly or after a pure lossy channel \cite{Phys.Rev.A.82.062316.2010}, and demonstrated experimentally \cite{Nat.Photonics.4.178.2010}.

Here, we propose a \emph{non-Gaussian} postselection method to emulate the photon subtraction used in coherent-state CV QKD protocols, which is employed right before the emission of the coherent states. One advantage of this \emph{virtual} photon subtraction is that, it can not only remove the complex physical operations, but also emulate the ideal photon-subtraction operations. Another advantage is that the postselection can be postponed after the parameter estimation, and therefore it can be adjusted flexibly to optimize the performance.
Besides, the postselection filter function does not need a cutoff amplitude as does the one in virtual NLA \cite{Phys.Rev.A.86.060302.2012,Phys.Rev.A.87.020303.2013,Nat.Photonics.8.333.2014}, because it is bounded.

Furthermore, our preliminary tests on the information reconciliation of the postselected non-Gaussian data suggest that the multidimensional reconciliation
algorithm can be directly used for the \emph{virtual} photon-subtraction method. Especially in the low signal-to-noise ratio regime, the performance of reconciliating the postselected non-Gaussian data is even better than that of the Gaussian data, which implies the feasibility of implementing this method practically.

This paper is organized as follows: In Sec. II, we introduce some basics of the photon subtraction and propose the equivalent postselection method of the photon-subtraction operation in a coherent-state CV QKD protocol. In Sec. III, we present the performance of the \emph{virtual} photon subtraction through numerical simulation, the optimal choice of the parameter in \emph{virtual} photon subtraction, and the tests of information reconciliation on the postselected non-Gaussian data. In Sec. IV, we summarize the paper.

%½²CVµÄ¾àÀëÊǸöÎÊÌâ
%
%½²NLA×÷Ϊ¾À²øÌá´¿¿ÉÒÔÌáÉý¾àÀë
%
%½²PSÒ²¿É¾À²øÌá´¿£¬ÉõÖÁ·Ç¸ß˹ҲÄÜÌáÉý¾àÀ룬µ«ÊÇÊÕµ½Êµ¼ÊÆ÷¼þµÄÓ°Ïì¡£
%
%½²ÎÒÃÇÌá³öµÄÕâ¸öµÄ˼·£¬·ÖÎöÁËʲô£¬Ð§¹û´óÖÂÈçºÎ
%
%ÎÄÕ½ṹ

\section{Photon Subtraction and its equivalent postselection in CV QKD}
Photon subtraction can enhance the entanglement of the TMSV state. Various scenarios were proposed, including applying photon subtraction on one mode or both modes of the TMSV \cite{Phys.Rev.A.73.042310.2006}, directly or after a pure lossy channel \cite{Phys.Rev.A.82.062316.2010}.
To make our derivation self-contained, in this section, we first introduce the basics of photon subtraction on a TMSV state (noted as the photon-subtracted TMSV state), then propose its equivalent non-Gaussian postselection in CV QKD protocols (noted as \emph{virtual} photon subtraction).

\subsection{Photon Subtraction of a TMSV State}
The TMSV state involves two modes $A$ and $B$. $\{ \hat a, {{\hat a}^\dag } \}$ and $\{ \hat b, {\hat b}^\dag \}$ denote the annihilation and creation operator of modes $A$ and $B$, respectively, where $[ {\hat a,{{\hat a}^\dag }} ] = [ {\hat b,{{\hat b}^\dag }} ] = 1$. A TMSV state can be expressed by
\begin{equation}
\left| {\left. {{\rm{TMSV}}} \right\rangle } \right.  = \sqrt {1 - {\lambda ^2}} \sum\limits_{n = 0}^\infty  {{\lambda ^n}\left| {\left. {n,n} \right\rangle } \right.},
\end{equation}
where $\lambda  \in \left[ {0,1} \right)$, $\left| {\left. {m,n} \right\rangle } \right. = {\left| {\left. m \right\rangle } \right._A} \otimes {\left| {\left. n \right\rangle } \right._B}$, and ${\left\{ {\left| {\left.n\right\rangle}\right.}\right\}_{n\in \mathbb{N}}}$ denotes the Fock state.

%Photon subtraction of a TMSV state could be operated on both modes $A$ and $B$ or only on one mode of them. It is shown that subtracting $k$ photons from mode $A$ equals to adding $k$ photons on mode $B$ up to a normalization factor [?Patron]. However, the post-selection method we proposed can only simulate the photon subtraction on one mode, thus in this paper we only consider the photon subtraction of one mode.

\begin{figure}[t]
  \includegraphics[width=3.45in]{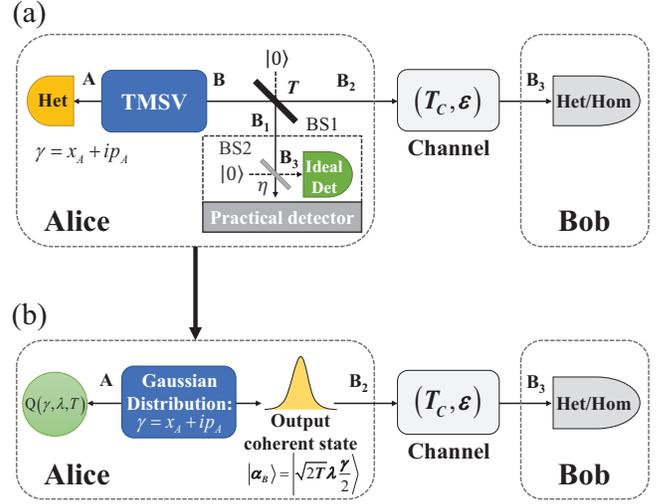}
  \caption{(Color online) (a) Entanglement-based (EB) scheme of CV QKD with photon subtraction. (b) Prepare-and-measure (PM) scheme of CV QKD with equivalent postselection as \emph{virtual} photon subtraction. Het: heterodyne detection; Hom: homodyne detection; BS1(2): beams plitter; $\gamma$: Alice's measurement result; $\lambda$: parameter of TMSV; $Q\left(\gamma,\lambda,T\right)$: postselection filter function; $T(\eta)$: transmittance of BS1(2); $T_C, \varepsilon$: channel parameters.
   }\label{schemetic}
\end{figure}

The EB scheme of CV QKD with photon subtraction inside Alice is shown in Fig. \ref{schemetic}(a). After generating the TMSV state, Alice uses a beams plitter (BS1), with transmittance $T$, to split the mode $B$ into modes $B_1$ and $B_2$, getting a tripartite state $\rho_{AB_1B_2}$,
\begin{equation}
{\rho _{A{B_1}{B_2}}} = {U_{BS}}\left[ {\left| \rm{TMSV} \right\rangle \left\langle \rm{TMSV} \right| \otimes \left| 0 \right\rangle \left\langle 0 \right|} \right]U_{BS}^\dag.
\end{equation}
Then $B_1$ will be measured by a positive operator-valued measure (POVM) measurement $\{ {{\hat \Pi _0},{\hat \Pi _1}} \}$, and the modes $A$ and $B_2$ will be kept only when the POVM element $\hat \Pi _1$ clicks. The kept state $\rho _{A{B_2}}^{{\hat \Pi _1}}$ is denoted as the photon-subtracted TMSV state,
\begin{equation}
\rho _{A{B_2}}^{{{\hat \Pi }_1}} = \frac{{{{\rm{tr}} _{{B_1}}}( {{\hat \Pi _1}{\rho _{A{B_1}{B_2}}}} )}}{{{{\rm{tr}}_{A{B_1}{B_2}}} ( {{\hat \Pi _1}{\rho _{A{B_1}{B_2}}}} )}}.
\end{equation}
where ${\rm{tr}}_{X}(\cdot)$ is the partial trace of a multimode quantum state, and $P^{{{\hat \Pi }_1}} = {{\rm{tr}}_{A{B_1}{B_2}}}( {{{\hat \Pi }_1}{\rho _{A{B_1}{B_2}}}} )$ is the success probability of $\hat\Pi_1$ clicks.

 Different $\hat \Pi_1$ will lead to different types of photon-subtraction. A general photon subtraction operation is subtracting $k$ photons, which refers to $\hat \Pi_1 = |k\rangle \langle k|$ and can be realized by a photon number resolving (PNR) detector \cite{Rev.Sci.Instrum.82.071101.2011}. And it is shown in \cite{Phys.Rev.A.86.012328.2012} that the entanglement will increase as more photons are subtracted.
 The photon subtraction can also be extended to the mixture of subtracting different $k$ photons, ${{\hat \Pi }_1}{\rm{ = }}\sum\nolimits_{k = 0}^\infty  {{c_k}\left| k \right\rangle \left\langle k \right|}  $ and ${c_k} \in \left\{ {0,1} \right\}$, among which ${{\hat \Pi }_{on}}{\rm{ = }}\sum\nolimits_{k = 1}^\infty  {\left| k \right\rangle \left\langle k \right|}  =  \mathbb{I}- \left| 0 \right\rangle \left\langle 0 \right|$ corresponds to an on-off detector. See Appendix \ref{k-photon subtraction} for more details about the above two examples.

\subsection{Virtual Photon Subtraction in CV QKD via Postselection}
We suppose Alice uses the photon-subtracted TMSV state as the source of a CV QKD system, and she will perform a heterodyne detection on mode $A$.
As shown in Fig. \ref{schemetic}(a), the measurement result of a single-photon detector represents either keeping this state (click) or not keeping this state (no click). Alice needs to record this extra data for each TMSV, and will reveal it to Bob after Bob measures the mode $B_3$.

According to the extemality of Gaussian quantum states \cite{Phys.Rev.Lett.96.080502.2006,Phys.Rev.Lett.97.190502.2006,Phys.Rev.Lett.97.190503.2006}, the secret key rate of the state $\rho_{AB_3}^{\hat \Pi _1}$ is no less than a Gaussian state $\rho_{AB_3}^{G}$ which has the same covariance matrix, where $K(\rho_{AB_3}^{\hat \Pi _1}) \ge K(\rho_{AB_3}^{G})$. Thus we will use $\rho_{AB_3}^{G}$ to derive the secret key rate. Besides, the success probability of Alice's POVM measurement $P^{{{\hat \Pi }_1}}$ should also be taken into account. Thus, for the reverse reconciliation, the lower bound of the asymptotic secret key rate under collective attack is
\begin{equation}\label{SKR}
K\left( {\rho _{A{B_3}}^G} \right) = {P^{{{\hat \Pi }_1}}}\left( {\beta{I^G}\left( {A:B} \right) - {S^G}\left( {E:B} \right)} \right),
\end{equation}
where $\beta$ is the reconciliation efficiency, $I^G{\left(A:B\right)}$ is the mutual information between Alice and Bob, $S^G\left(E:B\right)$ is the Holevo bound \cite{QCQC.Nielsen.2000} of the mutual information between Bob and Eve. The calculation method of $K$ is shown in the Appendix \ref{SKRcal}.

Next, we will present the equivalent \emph{virtual} photon subtraction via postselection according to Alice's measurement results, which will benefit the system. First, it is not necessary to accomplish the practical photon subtraction which reduces the complexity of the system. Second, it has better performance than the practical photon subtraction since one can emulate the ideal detector case.

The heterodyne detection and the POVM measurement $\{ {{\hat \Pi _0},{\hat \Pi _1}} \}$ are commutable since they are conducted on two different modes. Thus, Alice can perform the heterodyne detection on mode $A$ first, and then the POVM measurement on mode $B_1$. It is known that heterodyne detection on one mode of the TMSV state will project the other mode onto a coherent state; thus after BS1, the state of modes $B_1$ and $B_2$, given that Alice's heterodyne measurement results are $\left\{ {{x_A},{p_A}} \right\}$, is ${\left| {{\varphi ^{\left( {{x_A},{p_A}} \right)}}} \right\rangle _{{B_1}{B_2}}} = {\left| {\sqrt {1 - T} \alpha } \right\rangle _{{B_1}}}{\left| {\sqrt T \alpha } \right\rangle _{{B_2}}}$, where ${{\alpha  = \sqrt 2 \lambda \left( {{x_A} - i{p_A}} \right)} \mathord{\left/ {\vphantom {{\alpha  = \sqrt 2 \lambda \left( {{x_A} + i{p_A}} \right)} 2}} \right. \kern-\nulldelimiterspace} 2}$.
The success probability of subtracting $k$ photons, given Alice's heterodyne measurement results, will be the function of $\left\{ {{x_A},{p_A}} \right\}$,
\begin{equation}
\begin{array}{*{20}{l}}
{{P^{{{\hat \Pi }_1}}}\left( {k|{x_A},{p_A}} \right) = {{\left| {\left\langle {k}
 \mathrel{\left | {\vphantom {k {\sqrt {1 - T} \alpha }}}
 \right. \kern-\nulldelimiterspace}
 {{\sqrt {1 - T} \alpha }} \right\rangle } \right|}^2}}\\
{ = \exp \left[ { - \frac{{\left( {1 - T} \right){\lambda ^2}}}{2}\left( {x_A^2 + p_A^2} \right)} \right]{{ \cdot {{\left[ {\frac{{\left( {1 - T} \right){\lambda ^2}}}{2}\left( {x_A^2 + p_A^2} \right)} \right]}^k}} \mathord{\left/
 {\vphantom {{ \cdot {{\left[ {\frac{{\left( {1 - T} \right){\lambda ^2}}}{2}\left( {x_A^2 + p_A^2} \right)} \right]}^k}} {k!}}} \right.
 \kern-\nulldelimiterspace} {k!}}}
\end{array}
\end{equation}
Then the mixed state output from Alice's station will be
\begin{equation}\label{PS_EQ}
\rho _{{B_2}}^{\left(k\right)} = \int {\mathbf{d}{x_A}\mathbf{d}{p_A}\underbrace {\frac{{{P^{{{\hat \Pi }_1}}}\left( {k|{x_A},{p_A}} \right)}}{{{P^{{{\hat \Pi }_1}}}\left( k \right)}}}_{\scriptstyle{\rm{weighting~function}}}{P_{{x_A},{p_A}}}\left| {\sqrt T \alpha } \right\rangle \left\langle {\sqrt T \alpha } \right|},
\end{equation}
where ${P_{{x_A},{p_A}}} = \frac{1}{{\pi \left( {V + 1} \right)}}\exp \left( { - \frac{{x_A^2 + p_A^2}}{{V + 1}}} \right)$ is the Gaussian distribution of Alice's heterodyne measurement results, and $V = {{\left( {1 + {\lambda ^2}} \right)} \mathord{\left/
 {\vphantom {{\left( {1 + {\lambda ^2}} \right)} {\left( {1 - {\lambda ^2}} \right)}}} \right.
 \kern-\nulldelimiterspace} {\left( {1 - {\lambda ^2}} \right)}}$ is the variance of the TMSV state.

In the case where Alice does not use any kind of photon-subtraction operation, the output mixed state is
\begin{equation}
\rho _{{B}}^{\left(G\right)} = \int {\mathbf{d}{x_A}\mathbf{d}{p_A}{P_{{x_A},{p_A}}}\left| { \alpha } \right\rangle \left\langle { \alpha } \right|}.
\end{equation}
Compared with the postselected state in Eq. (\ref{PS_EQ}), there are two differences. Firstly, there is an additional weighting function in Eq. (\ref{PS_EQ}),
 \begin{equation}
 W = \frac{{{P^{{{\hat \Pi }_1}}}\left( {k|{x_A},{p_A}} \right)}}{{{P^{{{\hat \Pi }_1}}}\left( k \right)}},\nonumber
 \end{equation}
which leads to a filter function, or acceptance probability, of each pair of $\left\{ {{x_A},{p_A}} \right\}$,
\begin{equation}
 Q\left( {\gamma ,\lambda ,T} \right) = {{{P^{{{\hat \Pi }_1}}}\left( k \right)}}W= {{P^{{{\hat \Pi }_1}}}\left( {k|{x_A},{p_A}} \right)}.\label{Filter Function}
 \end{equation}
Second, the output coherent state needs to go through the BS1 with transmittance $T$, which can be emulated via generating a coherent state with a smaller mean value $\sqrt T \alpha$.

Thus, after exchanging Alice's heterodyne measurement and the photon-subtraction operation, we get the equivalent \emph{virtual} photon subtraction via postselection of Alice's measurement results and scaling the mean value of output coherent state by a factor $\sqrt T$. The postselection filter function is shown in Eq.(\ref{Filter Function}).

In summary, the PM scheme of CV QKD using \emph{virtual} photon subtraction, depicted in Fig. \ref{schemetic}(b), is as follow:

{\emph{Step 1.}} Alice generates a coherent state  $ \left| \alpha  \right\rangle $, where $\alpha  = \sqrt {2T} \lambda {\gamma  \mathord{\left/
 {\vphantom {\gamma  2}} \right.
 \kern-\nulldelimiterspace} 2}$, $\gamma  = {x_A} + i{p_A}$, and $\left\{{x_A}, {p_A}\right\}$ are chosen randomly from a Gaussian distributed set with zero mean and variance $V_{x_A} = V_{x_B} = {{\left( {V + 1} \right)} \mathord{\left/
 {\vphantom {{\left( {V + 1} \right)} 2}} \right. \kern-\nulldelimiterspace} 2}$. $T$ is the transmittance of BS1 and $V$ is the variance of the equivalent TMSV. Then she sends the coherent state to Bob.

{\emph{Step 2.}} After receiving the state, Bob will perform homodyne or heterodyne detection, and the measurement results are denoted by ${x_B, p_B}$.

{\emph{Step 3.}} \emph{Steps 1} and \emph{2} are repeated until they collect enough data. Alice will decide which data will be accepted, and reveals the decisions to Bob. The acceptance probability for each data is $Q\left( {\gamma ,\lambda ,T} \right)$ as in Eq. (\ref{Filter Function}). Then Alice and Bob use the accepted data to finish the postprocessing steps, including parameter estimation, information reconciliation and privacy amplification.

Since Alice reveals her decision of whether or not she accepts each data after Bob's measurement, the discarded states can be seen as the decoy states,
%is a widely used technique in discrete variable protocols \cite{Phys.Rev.Lett.91.057901.2003, Phys.Rev.Lett.94.230503.2005, Phys.Rev.Lett.94.230504.2005}
which are used in the former non-Gaussian protocol \cite{Phys.Rev.A.83.042312.2011} to enhance the security.

We note that for the practical implementation, the secret key rate, given by Eq. (\ref{SKR}), should be modified to take finite-size effects into account \cite{Phys.Rev.A.81.062343.2010}, which indicates that the acceptance probability will influence the finite-size analysis. However, this will not affect the effectiveness of our method; therefore, we only consider the asymptotic rate here for simplicity.

\section{Performance of the protocols}
In this section, we first present the performance of the protocols using photon subtraction in terms of secret key rate and tolerable excess noise, through numerical simulations. Then we discuss the transmittance of Alice's BS1, and the reconciliation efficiency of the postselected non-Gaussian data.
As described in Sec. II B, we will use Eq.(\ref{SKR}) as the asymptotic secret key rate. Since it only involves the Gaussian state, the covariance matrix of $\rho_{AB_3}^{\left(k\right)}$ will be sufficient to get the rate, which can be gotten according to Alice's and Bob's accepted data when implementing this protocol practically. Here in the simulation we assume that the channel can be characterized by two parameters: the channel transmittance $T_C$ and excess noise $\varepsilon$, which means if the covariance matrix of $\rho_{AB_2}^{\left(k\right)}$ is
\begin{equation}
\gamma _{A{B_2}}^{\left( k \right)} = \left( {\begin{array}{*{20}{c}}
{{V_A\mathbb{I}}}&{{\phi _{AB}}{\sigma _Z}}\\
{{\phi _{AB}}{\sigma _Z}}&{{V_B\mathbb{I}}}
\end{array}} \right),
\end{equation}
where $\mathbb{I} = $ diag(1,1) and $\sigma _Z =$ diag(1,-1), then, after the channel transmission,
\begin{equation}
\gamma _{A{B_3}}^{\left( k \right)} = \left( {\begin{array}{*{20}{c}}
{{V_A\mathbb{I}}}&{\sqrt {{T_C}} {\phi _{AB}}{\sigma _Z}}\\
{\sqrt {{T_C}} {\phi _{AB}}{\sigma _Z}}&{ {{T_C}\left( {{V_B}+ \chi } \right) } \mathbb{I}}
\end{array}} \right),
\end{equation}
where $\chi = {{{\left( {1 - {T_C}} \right)} \mathord{\left/
 {\vphantom {{\left( {1 - {T_C}} \right)} {{T_C}}}} \right.
 \kern-\nulldelimiterspace} {{T_C}}}}+\varepsilon$. The explicit form of $\gamma_{AB_2}^{\left(k\right)}$ can be found in Appendix \ref{CM of PS}. And for the rest of the paper, we assume Bob uses homodyne detection.
We note that Eve's optimal attack for this non-Gaussian protocol is still an open question.

\subsection{Secret Key Rate and Tolerable Excess Noise}
\begin{figure}[t]
\includegraphics[width=2.9in]{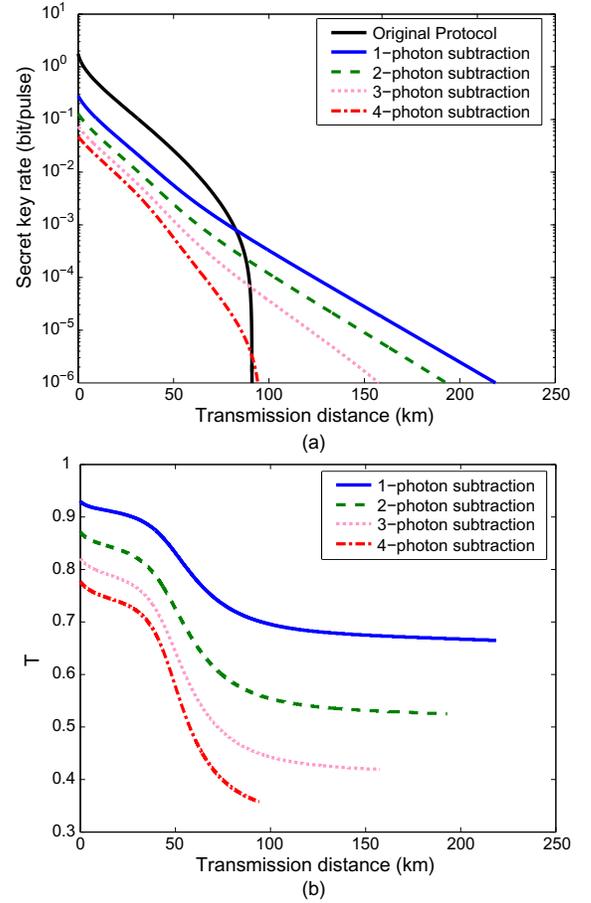}
%\subfigure[]{\includegraphics[width=3in]{03-2-kPS-Topt-ForBestSKR.eps}}
\caption{(Color online) (a) The maximal secret key rate at each transmission distance, when changing the transmittance $T$ of Alice's BS1. (b) The optimal $T$ for the maximal secret key rate in (a). The uppermost black solid line in (a) represents the case of original protocol. Other lines represent one-photon subtraction (blue solid line), two-photon subtraction (green dashed line), three-photon subtraction (pink dotted line), and four-photon subtraction (red dash-dotted line), respectively. The simulation parameters are as follows: the variance of TMSV state is $V = 20$, channel loss is $a = 0.2$dB/km, excess noise is $\varepsilon = 0.01$, and reconciliation efficiency is $\beta = 0.95$.
}\label{BestSKR}
\end{figure}

For the \emph{virtual} $k$-photon subtraction, the transmittance $T$ of Alice's BS1 can be chosen arbitrarily from 0 to 1, which will result in the change of overall acceptance probability ${P^{{{\hat \Pi }_1}}}\left( k \right)$ and also the covariance matrix of $\gamma _{A{B_2}}^{\left( k \right)}$. Thus, for each transmission distance, the secret key rate varies with different $T$, and there should exist an optimal choice of $T$ for each distance to maximize the secret key rate. Figure \ref{BestSKR}(a) shows the maximal secret key rate at each distance for all possible $T$. And Fig. \ref{BestSKR}(b) shows the optimal choice of $T$ for each distance, specifically, only for distances with secret key rate more than $10^{-6}$, respectively.

\begin{figure}[t]
\includegraphics[width=2.9in]{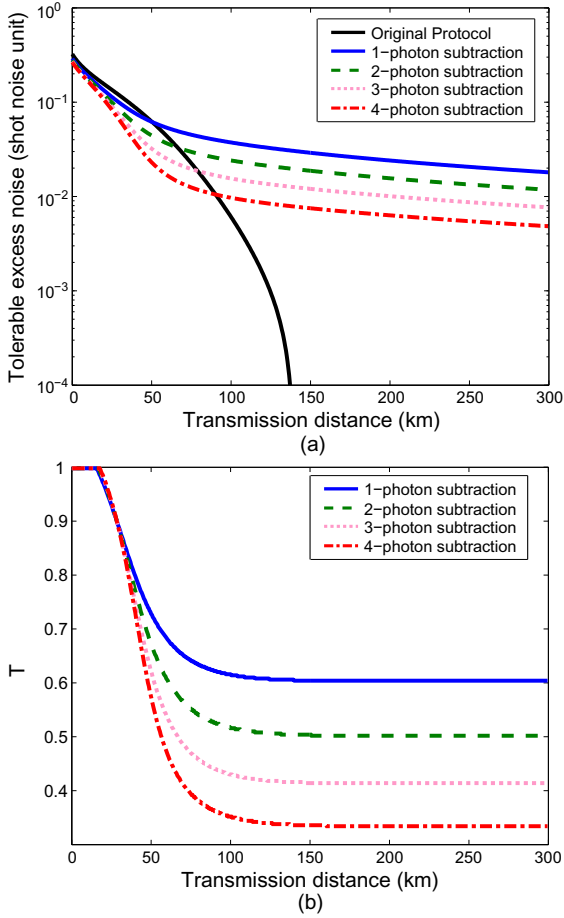}
%\subfigure[]{\includegraphics[width=3in]{04-2-kPS-Topt-forMaxEN}}
\caption{(Color online) (a) The maximal tolerable excess noise at each transmission distance, when changing the transmittance $T$ of Alice's BS1. (b) The optimal $T$ for the maximal tolerable excess noise in (a). The uppermost black solid line in (a) represents the case of original protocol. Other lines represent one-photon subtraction (blue solid line), two-photon subtraction (green dashed line), three-photon subtraction (pink dotted line), and four-photon subtraction (red dash-dotted line), respectively. The simulation parameters are $V = 20$, $a = 0.2$dB/km, $\varepsilon = 0.01$, and $\beta = 0.95$.
}\label{BestTEN}
\end{figure}

The black solid line in Fig. \ref{BestSKR}(a) represents the case of the original protocol, which is outperformed by the protocols of using photon subtraction at long-distance range, especially the case of using one-photon subtraction (blue solid line). This implies one of the advantages of using photon subtraction, that is, expanding the maximal transmission distance.
However, for the short-distance range, even for the optimal choice of $T$, the secret key rate is still worse than the original protocol. One reason is the limited acceptance probability, which is below 0.25 under the parameters we used here (see Fig. \ref{psucc}).

Fig. \ref{BestSKR}(a) also shows that the one-photon subtraction has the longest transmission distance compared to the other cases.
The main reason is that, when subtracting more photons, the non-Gaussianity is higher \cite{Phys.Rev.A.86.012328.2012}, which means the
Gaussian state that has the same covariance matrix, $\rho_{AB_2}^G$, is more noisy. Thus, when employing the extremality of Gaussian quantum states to do the security analysis, it gets a
worse result. More specifically, we rewrite the covariance matrix $\gamma _{A{B_2}}^{\left( k \right)}$ as follows:
\begin{equation}
\gamma _{A{B_2}}^{\left( k \right)} = \left( {\begin{array}{*{20}{c}}
{{V_A\mathbb{I}}}&{\sqrt {{\eta _A}\left( {V_A^2 - 1} \right)} {\sigma _Z}}\\
{\sqrt {{\eta _A}\left( {V_A^2 - 1} \right)} {\sigma _Z}}&{{\eta _A}\left( {{V_A} + \chi_A } \right)\mathbb{I}}
\end{array}} \right),
\end{equation}
where ${V_A} = 2\tilde V - 1$, $\chi_A = {{{\left( {1 - {\eta _A}} \right)} \mathord{\left/
 {\vphantom {{\left( {1 - {\eta _A}} \right)} {{\eta _A}}}} \right.
 \kern-\nulldelimiterspace} {{\eta _A}}}}$, and
\begin{equation}
 {\eta _A} = {\lambda ^2}T\frac{{k + 1}}{{k + {\lambda ^2}T}}.
\end{equation}
This means, in the viewpoint of calculating the secret key rate, the $k$-photon-subtracted TMSV can be seen as a source with an extra loss on the $B_2$ mode before being emitted into the channel.
And $\eta_A$ decreases as the photon-subtraction number $k$ increases, especially when the optimal $T$ is lower for higher $k$ at long-distance range, which implies that subtracting more photons will result in more loss. Therefore, subtracting one photon shows better performance under the conditions of Fig. \ref{BestSKR}, i.e., the initial variance of TMSV is 20.

Tolerable excess noise represents another aspect of a protocol.
Figure \ref{BestTEN}(a) shows the maximal tolerable excess noise at each distance for all possible $T$, and Fig. \ref{BestTEN}(b) shows the optimal choice of $T$ for each distance. Similar to Fig. \ref{BestSKR}(a), the case of original protocol (black solid line) is also outperformed by other cases at long transmission distance range, which implies another advantage of using photon subtraction that increases the maximal tolerable excess noise for distant users. It is also shown that when the channel is less noisy, for instance, $\varepsilon  \sim 0.005$, all four photon-subtraction operations will expand the maximal transmission distance to more than $200\rm{km}$.
On the other hand, one could notice that the optimal choice of $T$ for the maximal tolerable excess noise shows a different form from the one for the maximal secret key rate. That is because the tolerable excess noise is not affected by the overall acceptance probability, while the secret key rate is.

\subsection{The Transmittance of Alice's Beamsplitter}
In \emph{Step 1} of the PM scheme using \emph{virtual} photon subtraction described in Sec. II, it requires Alice to know the value of $T$ in advance. For a relatively stable system and environment, Alice can use the data of the last run to approximately estimate the optimal $T$ for this run. However, when the system or environment changes rapidly, the above method may not result in a suitable estimation. In this case, one can linearly scale Alice's heterodyne measurement results first and then accomplish the postselection.
After the linearly scaling, Alice's data follows a new Gaussian distribution $P{'_{{X_A},{P_A}}}$ with a different variance ${V}'$, which can be regarded as the heterodyne measurement results of a new equivalent TMSV. We assume
 $T_0$ is the estimated value according to the data of the last run, and $\eta$ is the real optimal choice of this run. Let ${X_A} = G{x_A}$,
${P_A} = G{p_A}$,
 and ${\alpha _A} = {{\sqrt 2 \lambda '\left( {{X_A} - i{P_A}} \right)} \mathord{\left/ {\vphantom {{\sqrt 2 \lambda '\left( {{X_A} - i{P_A}} \right)} 2}} \right. \kern-\nulldelimiterspace} 2} = \sqrt {{{{T_0}} \mathord{\left/
 {\vphantom {{{T_0}} \eta }} \right. \kern-\nulldelimiterspace} \eta }} \cdot\alpha  $, where $G = {{\sqrt {{T_0}} \lambda } \mathord{\left/
 {\vphantom {{\sqrt {{T_0}} \lambda } {\sqrt \eta  \lambda '}}} \right.
 \kern-\nulldelimiterspace} {\sqrt \eta  \lambda '}}$ and $\lambda {'^2} = {{\left( {V' - 1} \right)} \mathord{\left/
 {\vphantom {{\left( {V' - 1} \right)} {\left( {V' + 1} \right)}}} \right.
 \kern-\nulldelimiterspace} {\left( {V' + 1} \right)}}$. Then the state Alice initially sends out can be rewritten as
\begin{equation}
\begin{array}{l}
\rho _{{B_2}}^{\left( k \right)} = \int {\mathbf{d}{x_A}\mathbf{d}{p_A} {P_{{x_A},{p_A}}}\left| {\sqrt {{T_0}} \alpha } \right\rangle \langle \sqrt {{T_0}} \alpha |} \\
 = \int {\mathbf{d}{X_A}\mathbf{d}{P_A}  P{'_{{X_A},{P_A}}}\left| {\sqrt \eta  {\alpha _A}} \right\rangle \langle \sqrt \eta  {\alpha _A}|}
\end{array},
\end{equation}
where $P{'_{{X_A},{P_A}}} = {{{P_{{x_A},{p_A}}}} \mathord{\left/
 {\vphantom {{{P_{{x_A},{p_A}}}} {{G^2}}}} \right.
 \kern-\nulldelimiterspace} {{G^2}}}$, with $x_A$ and $p_A$ substituted by  ${{{X_A}} \mathord{\left/
 {\vphantom {{{X_A}} G}} \right. \kern-\nulldelimiterspace} G}$ and ${{{P_A}} \mathord{\left/
 {\vphantom {{{P_A}} G}} \right. \kern-\nulldelimiterspace} G}$,
is the new Gaussian distribution of the scaled data. Considering the explicit form of $P{'_{{X_A},{P_A}}}$, the variance $V'$ fulfills that $V'-1 = {{{T_0}} \mathord{\left/ {\vphantom {{{T_0}} \eta }} \right. \kern-\nulldelimiterspace} \eta } \cdot \left( {V - 1} \right)$.

After this linearly scaling step, Alice could do the equivalent postselection according to the new data ${{{X_A},{P_A}}}$. By traversing all possible value of $\eta$, one could find the optimal value of it to optimize the secret key rate at a certain distance. In this way, the key parameter of \emph{virtual} photon subtraction, the transmittance of Alice's BS1, can be adjusted after Alice and Bob estimate the channel parameters, which makes this method more flexible.

\begin{figure}[t]
\includegraphics[width=3in]{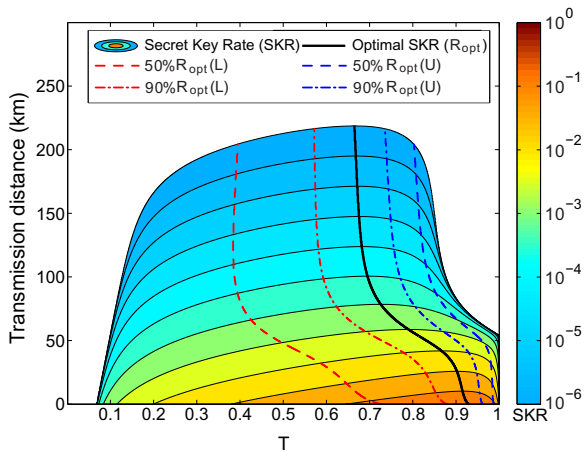}
\caption{(Color online) Secret key rate vs transmission distance and transmittance $T$ of Alice's BS1. The middle black solid line represents the optimal $T$ of each distance, when the secret key rate approaches its optimal value ($R_{opt}$). The left red dash-dotted (dashed) line represents the lower bound of $T$ for each distance, when its secret key rate is 90\% (50\%) of its optimum at that distance. The right blue dash-dotted (dashed) line represents the upper bound of $T$ for each distance, when its secret key rate is 90\% (50\%) of its optimum at that distance. The simulation parameters are: $V = 20$, $a = 0.2$dB/km, $\varepsilon = 0.01$, and $\beta = 0.95$.
}\label{SKR-T-TD}
\end{figure}

%Since $T$ is estimated only by the random sampled data, it may deviate from the optimal value for the whole set of data. Fortunately, this kind of deviation only decrease the performance slightly. Because as shown in
%Fig. \ref{SKR-T-TD}, the secret key rate can maintain higher than 90\% of its optimum (black solid line) at each distance, if the estimated $T$ is in the area between the left red dash-dotted line and the right blue dash-dotted line, which is more than 0.1. Usually speaking, this is larger than the possible statistical deviation, when the data block size is more than we take 10\%.

Another problem about the choice of $T$ is that if the secret key rate varies significantly with $T$ around its optimal value, then we need to estimate it accurately to maintain a relatively high performance, which requires complicated implementations in a practical system. Fortunately, as shown in Fig. \ref{SKR-T-TD}, the secret key rate varies slowly with $T$ at each distance around its optimal value (black solid line). Specifically, the secret key rate can maintain higher than 90\% of its optimal value ($R_{opt}$), if the estimated $T$ is in the area between the left red dash-dotted line and the right blue dash-dotted line. If one only requires that the secret key rate is higher than one-half of its optimum, then the choice of $T$ is much more flexible, i.e., within the area between the left red dashed line and the right dashed line.

\subsection{Reconciliation Efficiency of non-Gaussian Data}
In all of the above discussions, we assume the reconciliation efficiency $\beta$ is 0.95, which is a pretty good efficiency approached in the Gaussian case \cite{Nat.Photonics.7.378.2013}. Therefore, another thing one may be concerned about is, for the non-Gaussian data generated by \emph{virtual} photon subtraction, whether or not the information reconciliation still remains a relatively high efficiency as for the Gaussian data. We have carried out a preliminary test on the performance of the multidimensional reconciliation method, proposed in \cite{Phys.Rev.A.77.042325.2008}, on both using one-photon subtraction and not using any photon-subtraction cases (non-Gaussian and Gaussian cases, respectively) with simulated data, assuming the binary input additive white Gaussian noise channel (BIAWGNC).

The two sparse check matrices used here are of two different rates: one is 0.1 and another is 0.02. For each signal-to-noise ratio (SNR), we have tested more than 40 data blocks. Each data block contains $2^{20}$ bits (1~Mbits). Therefore, the number of successfully decoded data blocks and the average iteration number when the decoding process succeeds will represent the performance of the reconciliation method.

Table \ref{NGperformance} shows our test results in which, for the 0.1 rate matrix, non-Gaussian cases show a better performance than the Gaussian cases, on both the numbers of successfully decoded blocks and the average iteration number, given the same SNR. For the 0.02 rate matrix, the non-Gaussian cases show similar numbers of successfully decoded data blocks but less average iteration numbers compared to the Gaussian cases, except the last row. In the last row, the successful decoding probability of the Gaussian case drops significantly, while the non-Gaussian case only drops a little, provided there is more average iteration.

\begin{table}[t]
\begin{center}
\begin{tabular}{lccccc}\hline
      & SNR     & $\beta$& Type   & S/T & AIN  \\ \hline
$R$=0.1 & 0.1626	& 92.02\%& Gaussian      &	39/40	   &~103~ \\ %\cline{4-6}
	  &		    &	     & Non-Gaussian	 &  40/40	   & 82 \\ \cline{2-6}
	  & 0.1613	& 92.71\%& Gaussian      &	33/40	   & 134 \\ %\cline{4-6}
	  &	        & 	     & Non-Gaussian	 &  40/40	   & 102 \\ \cline{2-6}
	  & 0.1600	& 93.40\%& Gaussian      &	20/40	   & 151 \\ %\cline{4-6}
	  &		    &	     & Non-Gaussian	 &  34/40	   & 130 \\ \hline
     % & SNR     & Type   & Success/Total & AIN   & $\beta$ \\ \hline
$R$=0.02& 0.0301	& 93.37\%& Gaussian     &	47/48	   & 111 \\ %\cline{4-6}
	  &		    &	     & Non-Gaussian	 &  47/48	   & 101 \\ \cline{2-6}
	  & 0.0296	& 94.97\%& Gaussian      &	37/48	   & 190 \\ %\cline{4-6}
	  &	        & 	     & Non-Gaussian	 &  37/48	   & 174 \\ \cline{2-6}
	  & 0.0293	& 95.94\%& Gaussian      &	18/48	   & 157 \\ %\cline{4-6}
	  &		    &	     & Non-Gaussian	 &  33/48	   & 178 \\ \hline
\end{tabular}\caption{Performance comparison of the multidimensional reconciliation method between Gaussian and non-Gaussian data. R: the rate of sparse check matrix; SNR: signal-to-noise ratio; $\beta$: reconciliation efficiency; Type: the type of tested data; S/T: the number of successfully decoded data blocks/the number of total tested data blocks; AIN: average iteration number when the decoding process succeeds. }\label{NGperformance}
\end{center}
\end{table}	

In short, from our test, when considering a high reconciliation efficiency $\approx0.96$, the 0.02 rate check matrix shows a relatively high successful decoding probability, i.e., more than 60\%.
Besides, the two sparse check matrices were initially designed for Gaussian data, not specially designed for the photon-subtraction case. Thus, this result suggests that one may directly use the multidimensional reconciliation codes for photon subtraction. If the check matrix is specially designed for the non-Gaussian data, the reconciliation efficiency may be even higher.

\section{Conclusion}
In this paper, we proposed the \emph{virtual} photon-subtraction method in coherent-state CV QKD protocols, which can be accomplished by non-Gaussian postselection according to Alice's data. It can not only remove the complex physical operations, but also emulate the ideal operations which optimizes the performance of the CV QKD system. The main parameter, i.e., the transmittance of Alice's BS1, of this postselection method can be adjusted flexibly according to the channel parameters to optimize the secret key rate or tolerable excess noise at a certain distance. The numerical simulation shows that by choosing the optimal transmittance of Alice's BS1, the use of virtual photon subtraction will outperform the original protocol at long-distance regime.

Furthermore, our preliminary tests about the information reconciliation suggest that by using the multidimensional reconciliation
algorithm, the performance of reconciliating the postselected non-Gaussian data is even better than that of the Gaussian data. Specifically, for each SNR, either the successfully decoded blocks are higher or the average iteration numbers are lower which saves decoding time. In our tests, the two sparse check matrices were initially designed for Gaussian data, not for the non-Gaussian case, which suggests that one can directly use the multidimensional reconciliation method here. This implies the feasibility of implementing this \emph{virtual} photon-subtraction method practically.

\section*{Acknowledgement}
We would like to thank R. G. Patr\'{o}n and F. Grosshans for the helpful discussions. This work is supported by the National Science Fund for Distinguished Young Scholars of China (Grant No. 61225003), the State Key Project of National Natural Science Foundation of China (Grant No. 61531003), National Natural Science Foundation of China (Grant No. 61501414), and the National Hi-Tech Research and Development (863) Program.

\appendix

\section{Influence of the imperfect single-photon detector}\label{imperfect}

%In the following, we give the detailed description of CV QKD protocol when Alice uses photon subtraction first, and then show how the SPD's imperfections will reduce the performance by numerical simulation (Fig. \ref{sim1}).

%{\emph{Step 1.}} Alice generates one TMSV state with variance $V$, and keeps mode $A$ while sending mode $B$ passing through a BS with transmittance $T$, getting modes $B_1$ and $B_2$. Alice performs POVM measurement $\{ {{\hat \Pi _0},{\hat \Pi _1}} \}$ on mode $B_1$, and the measurement result is denoted by $A_{POVM}$. The mode $B_2$ will be sent to Bob through quantum channel.
%
%{\emph{Step 2.}} After Bob receiving the state, Bob will perform homodyne or heterodyne detection, and the measurement results are denoted by ${x_B, p_B}$. Alice will perform heterodyne detection on mode $A$, and the measurement results are denoted by ${x_A, p_A}$.
%
%{\emph{Step 3.}} Repeating {\emph{Step 1}} and {\emph{Step 2}} until they collect enough data. Alice will decide whether keeping or dropping each round according to $A_{POVM}$s, and reveal these decisions to Bob through public channel. Alice and Bob use the kept data to finish the post-processing steps, including parameter estimation, information reconciliation and privacy amplification.

\begin{figure}[t]
\includegraphics[width=3in]{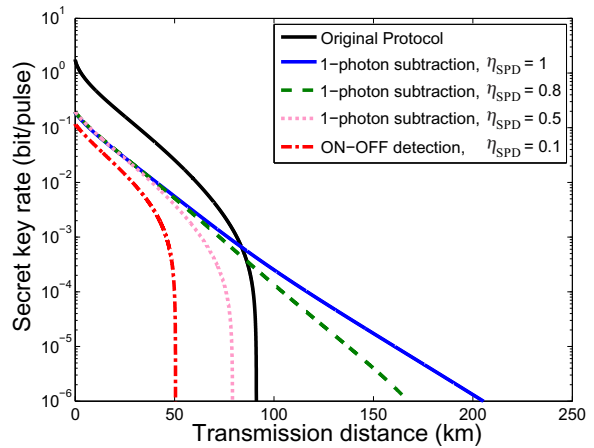}
\caption{(Color online) The detection efficiency of SPD will influence the performance of the scheme using photon subtraction. The lines from top to bottom are as follows: the original protocol without photon subtraction (black solid line),  with one-photon subtraction under unit DE (blue solid line), 0.8 DE (green dashed line), 0.5 DE (pink dotted line), and with on-off detector under 0.1 DE (red dash-dotted line), respectively. The simulation parameters are the variance $V = 20$, channel loss $a = 0.2$dB/km, excess noise $\varepsilon = 0.01$, the transmittance of BS1 $T = 0.8$, and reconciliation efficiency $\beta = 0.95$.
}\label{sim1}
\end{figure}

The perfect one-photon subtraction will require an ideal PNR detector. However, a practical PNR detector has imperfections, such as finite detection efficiency (DE) and dark count, which will reduce the maximal transmission distance. Because the average photon number for a TMSV state used in CV QKD is usually several tens, which means that the number of the legitimate count of a well-demonstrated PNR is much greater than the dark count, the finite detection efficiency ($\eta_d$) is the most significant factor. It is similar for the on-off detector based on the avalanched photodiode (APD). Thus, we only consider the effect of the limited detection efficiency.
As depicted in Fig. \ref{schemetic}(a), an extra beam splitter (BS2) with transmittance $1-\eta_d$ is put in front of an ideal detector to model the practical detector's finite detection efficiency. Figure \ref{sim1} shows how the SPD's nonunit detection efficiency will reduce the performance by numerical simulation. When the detection efficiency descends to 0.8 (green dashed line), although it still outperforms the original protocol, the maximal transmission distance decreases significantly. If the detection efficiency descends to 0.5 (pink dotted line), although it is achievable using superconducting transition-edge sensors at telecom wavelength \cite{Rev.Sci.Instrum.82.071101.2011}, the maximal transmission distance is worse than the original protocol. And it is even worse if one uses the commercial on-off SPD with only 0.1 detection efficiency based on APD (red dash-dotted line).

\section{Subtracting $k$ photons} \label{k-photon subtraction}
Here we use the same notation as depicted in Sec. II. After the BS1, the state is ${\rho _{A{B_1}{B_2}}} = \left| \psi  \right\rangle \left\langle \psi  \right|$, where

\begin{equation}
\begin{array}{l}
\left| \psi  \right\rangle  = {U_{BS}}\left| {\rm{TMSV}} \right\rangle  \otimes \left| 0 \right\rangle \\
 = \sqrt {1 - {\lambda ^2}} \sum\limits_{n = 0}^\infty  {{\lambda ^n}\left( {{U_{BS}}\left| {\left. {n,0} \right\rangle } \right.} \right)}  \otimes {\left| n \right\rangle _A}\\
 = \sqrt {1 - {\lambda ^2}} \sum\limits_{n = 0}^\infty  {\sum\limits_{l = 0}^n {{\lambda ^n}\sqrt {C_n^l{T^{n - l}}{{\left( {1 - T} \right)}^l}} } {{\left| {n,l,n - l} \right\rangle }_{A{B_1}{B_2}}}}
\end{array}
\end{equation}

\noindent and ${C_n^l}$ is the combinatorial number.

\begin{figure}[t]
\includegraphics[width=3in]{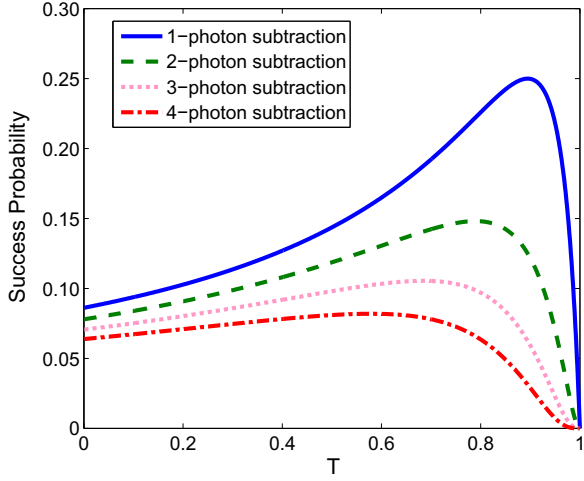}
\caption{(Color online) The success probability of subtracting $k$ photons of a TMSV with different transmittances $T$ of Alice's BS. The lines from top to bottom represent one-photon subtraction (blue solid line), two-photon subtraction (green dashed line), three-photon subtraction (pink dotted line), and four-photon subtraction (red dash-dotted line), respectively. The variance of TMSV is $V=20$, and Alice uses ideal SPD.
}\label{psucc}
\end{figure}

The success probability of $\hat \Pi_1 = |k\rangle \langle k|$ clicks on mode $B_1$ is
\begin{equation}
\begin{array}{l}
{P^{{{\hat \Pi }_1}}}\left( k \right) = t{r_{A{B_1}{B_2}}}\left( {{{\hat \Pi }_1}{\rho _{A{B_1}{B_2}}}} \right)\\
 = \left( {1 - {\lambda ^2}} \right)\sum\limits_{n = k}^\infty  {{\lambda ^{2n}}C_n^k{T^{n - k}}{{\left( {1 - T} \right)}^k}} \\
 = \left( {1 - {\lambda ^2}} \right){\left( {\frac{{1 - T}}{T}} \right)^k}\sum\limits_{n = k}^\infty  {{{\left( {{\lambda ^2}T} \right)}^n}C_n^k} \\
{\rm{ = }}\frac{{1 - {\lambda ^2}}}{{1 - T{\lambda ^2}}}{\left[ {\frac{{{\lambda ^2}\left( {1 - T} \right)}}{{1 - T{\lambda ^2}}}} \right]^k}
\end{array}.
\end{equation}
And its relationship with the transmittance of Alice's BS1 is shown in Fig. \ref{psucc}.

Then the $k$-photon subtracted state is
 $$\rho _{A{B_2}}^{\left(k\right)} = \frac{{\left\langle k \right|{\rho _{A{B_1}{B_2}}}\left| k \right\rangle }}{{{P^{{{\hat \Pi }_1}}}\left( k \right)}} = \left| {{\zeta ^{\left( k \right)}}} \right\rangle \left\langle {{\zeta ^{\left( k \right)}}} \right|,$$ where
 \begin{equation}
 \left| {{\zeta ^{\left( k \right)}}} \right\rangle  = \frac{{\left\langle {k}
 \mathrel{\left | {\vphantom {k \psi }}
 \right. \kern-\nulldelimiterspace}
 {\psi } \right\rangle }}{{\sqrt {{P^{{{\hat \Pi }_1}}}\left( k \right)} }} = \sum\limits_{n = k}^\infty  {\sqrt {p_n^{\left( k \right)}} {{\left| {n,n - k} \right\rangle }_{A{B_2}}}},
 \end{equation}
and
\begin{equation}
p_n^{\left( k \right)} = \frac{{{\lambda ^{2n}}C_n^k{T^n}}}{{\sum\limits_{n = k}^\infty  {{{\left( {{\lambda ^2}T} \right)}^n}C_n^k} }}.
\end{equation}

As a comparison, for the on-off detector, the success probability is
\begin{equation}
\begin{array}{l}
{P^{{{\hat \Pi }_1}}}\left( {on} \right) = 1 - {P^{{{\hat \Pi }_1}}}\left( {k = 0} \right)\\
 = 1 - \left( {1 - {\lambda ^2}} \right)\sum\limits_{n = 0}^\infty  {{{\left( {{\lambda ^2}T} \right)}^n}}  = \frac{{\left( {1 - T} \right){\lambda ^2}}}{{1 - {\lambda ^2}T}}
\end{array}.
\end{equation}
The final state is a mixed state, such that
\begin{equation}
\rho _{A{B_2}}^{on - off} = \sum\limits_{k = 1}^\infty  {\frac{{{P^{{{\hat \Pi }_1}}}\left( k \right)}}{{{P^{{{\hat \Pi }_1}}}\left( {on} \right)}}} \rho _{A{B_2}}^{\left(k\right)}.
\end{equation}

\section{Calculation of the Secret Key Rate} \label{SKRcal}
Suppose the final state $\rho_{AB}^G$ is a Gaussian state with covariance matrix
\begin{equation}
\gamma _{A{B}}^G = \left( {\begin{array}{*{20}{c}}
{{V_1}\mathbb{I}}&{\phi {\sigma _Z}}\\
{\phi {\sigma _Z}}&{{V_2}\mathbb{I}}
\end{array}} \right),
\end{equation}
where $\mathbb{I}$ is diag(1,1), and $\sigma _Z$ is diag(1,-1), and Alice always uses heterodyne detection. The secret key rate of reverse reconciliation is
\begin{equation}
{{K^{{\rm{Hom}}}} = \beta {I^{{\rm{Hom}}}}\left( {A:{B}} \right) - {S^{{\rm{Hom}}}}\left( {E:{B}} \right)},
\end{equation}
where the superscript Hom means Bob using homodyne detection, and $\beta$ is the reconciliation efficiency.

Therefore,
\begin{equation}
I^{\rm{Hom}}\left( {A:B} \right) = \frac{1}{2}{\log _2}\frac{{{V_A}}}{{V_{A|B}^{{\rm{Hom}}}}},
\end{equation}
where ${V_A} = {{\left( {{V_1} + 1} \right)} \mathord{\left/ {\vphantom {{\left( {{V_1} + 1} \right)} 2}} \right.
 \kern-\nulldelimiterspace} 2}$, $V_B = V_2$, and
\begin{equation}
V_{A|B}^{{\rm{Hom}}} = {V_A} - \frac{{{\phi ^2}}}{{2{V_B}}} = \frac{{{V_1} + 1}}{2} - \frac{{{\phi ^2}}}{{2{V_2}}}.
\end{equation}

%${V_A} = {{\left( {{V_1} + 1} \right)} \mathord{\left/ {\vphantom {{\left( {{V_1} + 1} \right)} 2}} \right.
% \kern-\nulldelimiterspace} 2}$, $V_B = V_2$, and $V_{A|B}^{{\rm{Hom}}} = {V_A} - {{{\phi ^2}} \mathord{\left/
% {\vphantom {{{\phi ^2}} {\left( {2{V_B}} \right)}}} \right.
% \kern-\nulldelimiterspace} {\left( {2{V_B}} \right)}} = {{\left( {{V_1} + 1} \right)} \mathord{\left/
% {\vphantom {{\left( {{V_1} + 1} \right)} 2}} \right.
% \kern-\nulldelimiterspace} 2} - {{{\phi ^2}} \mathord{\left/
% {\vphantom {{{\phi ^2}} {\left( {2{V_2}} \right)}}} \right.
% \kern-\nulldelimiterspace} {\left( {2{V_2}} \right)}}$.

By assuming Eve can purify the whole system, $S\left( {E:{B}} \right) = S\left( E \right) - S\left( {E|{B}} \right) = S\left( {A{B}} \right) - S\left( {A|{B}} \right)$. And $S\left(A{B}\right)$ is a function of the symplectic eigenvalues ${\lambda _{1,2}}$ of $\gamma_{AB}^G$, which is
\begin{equation}
S\left( {AB} \right) = G\left[ {{{\left( {{\lambda _1} - 1} \right)} \mathord{\left/
 {\vphantom {{\left( {{\lambda _1} - 1} \right)} 2}} \right.
 \kern-\nulldelimiterspace} 2}} \right] + G\left[ {{{\left( {{\lambda _2} - 1} \right)} \mathord{\left/
 {\vphantom {{\left( {{\lambda _2} - 1} \right)} 2}} \right.
 \kern-\nulldelimiterspace} 2}} \right],
\end{equation}
where
\begin{equation}
G\left( x \right) = \left( {x + 1} \right){\log _2}\left( {x + 1} \right) - x{\log _2}x,
\end{equation}
and
\begin{equation}
\lambda _{1,2}^2 = \frac{1}{2}\left[ {\Delta  \pm \sqrt {{\Delta ^2} - 4{D^2}} } \right],
\end{equation}
where we have used the notations
\begin{equation}
\begin{array}{l}
\Delta  = V_1^2 + V_2^2 - 2{\phi ^2},\\
D = {V_1}{V_2} - {\phi ^2}.
\end{array}
\end{equation}
And $S^{\rm{Hom}}\left(A|B\right) = G\left[ {{{\left( {{\lambda _3} - 1} \right)} \mathord{\left/
 {\vphantom {{\left( {{\lambda _3} - 1} \right)} 2}} \right. \kern-\nulldelimiterspace} 2}} \right]$ is a function of the symplectic eigenvalue $\lambda_3$ of the covariance matrix $\gamma_A^{b}$ of the $A$ mode after Bob's homodyne detection, where ${\lambda _3} = \sqrt {{V_1}\left( {{V_1} - {{{\phi ^2}} \mathord{\left/ {\vphantom {{{\phi ^2}} {{V_2}}}} \right. \kern-\nulldelimiterspace} {{V_2}}}} \right)} $. Thus, the secret key rate when Bob using homodyne detection is
\begin{equation}
{K^{{\rm{Hom}}}} = {I^{{\rm{Hom}}}}\left( {A:B} \right) - \left[ {S\left( {AB} \right) - {S^{{\rm{Hom}}}}\left( {A|B} \right)} \right].
\end{equation}

%While for Bob using heterodyne detection, the $I(A:B)$ is
%\begin{equation}
%I^{\rm{Het}}\left( {A:B} \right) = {\log _2}\frac{{{V_A}}}{{V_{A|B}^{{\rm{Het}}}}},
%\end{equation}
%where ${V_A} = {{\left( {{V_1} + 1} \right)} \mathord{\left/
% {\vphantom {{\left( {{V_1} + 1} \right)} 2}} \right.
% \kern-\nulldelimiterspace} 2}$, ${V_B} = {{\left( {{V_2} + 1} \right)} \mathord{\left/
% {\vphantom {{\left( {{V_2} + 1} \right)} 2}} \right.
% \kern-\nulldelimiterspace} 2}$,
% and
% \begin{equation}
%V_{A|B}^{{\rm{Het}}} = {V_A} - \frac{{{\phi ^2}}}{{4{V_B}}} = \frac{{{V_1} + 1}}{2} - \frac{{{\phi ^2}}}{{2\left( {{V_2} + 1} \right)}}.
% \end{equation}
%${S\left( {AB} \right)}$ is the same as the above homodyne case, while ${S^{{\rm{Het}}}}\left( {A|B} \right) = G\left[ {{{\left( {{\lambda _3}' - 1} \right)} \mathord{\left/ {\vphantom {{\left( {{\lambda _3}' - 1} \right)} 2}} \right. \kern-\nulldelimiterspace} 2}} \right]$ where ${\lambda _3}' = {V_1} - {{{\phi ^2}} \mathord{\left/ {\vphantom {{{\phi ^2}} {\left( {{V_2} + 1} \right)}}} \right. \kern-\nulldelimiterspace} {\left( {{V_2} + 1} \right)}}$. Thus, the secret key rate when Bob using heterodyne detection is
%\begin{equation}
%{R^{{\rm{Het}}}} = {I^{{\rm{Het}}}}\left( {A:B} \right) - \left[ {S\left( {AB} \right) - {S^{{\rm{Het}}}}\left( {A|B} \right)} \right].
%\end{equation}

\section{Covariance matrix of $k$-photon subtracted TMSV state} \label{CM of PS}
Suppose $\gamma^{\left(k\right)}$ represents the covariance matrix of $\rho_{AB_2}^{\left(k\right)}$, % whose element is defined as
%\begin{equation}
%{\gamma _{ij}^{\left(k\right)}}: = \frac{1}{2}\left\langle {\left\{ {\Delta {{\hat r}_i},\Delta {{\hat r}_j}} \right\}} \right\rangle,
%\end{equation}
%where $\Delta {{\hat r}_i}: = {{\hat r}_i} - \left\langle {{{\hat r}_i}} \right\rangle$, ${\bf{\hat r}}: = {\left( {{{\hat x}_A},{{\hat p}_A},{{\hat x}_B},{{\hat p}_B}} \right)^T}$ is the vector of quadratures, and $\left\{ , \right\}$ is the anticommutator [].
and it has the following formula,
\begin{equation}
{\gamma ^{\left( k \right)}} = \left( {\begin{array}{*{20}{c}}
{\left\langle {\hat x_A^2} \right\rangle \mathbb{I}}&{\left\langle {{{\hat x}_A}{{\hat x}_B}} \right\rangle {\sigma _Z}}\\
{\left\langle {{{\hat x}_A}{{\hat x}_B}} \right\rangle {\sigma _Z}}&{\left\langle {\hat x_B^2} \right\rangle \mathbb{I}}
\end{array}} \right).
\end{equation}

Suppose $x', p'$ are the heterodyne measurement results of mode $A$, and $x$ is the homodyne measurement result of mode $B$. Then,
\begin{equation}
\begin{array}{l}
\left\langle {\hat x_A^2} \right\rangle  = 2 \cdot {\int {{{x'}^2} P\left( {x',p',x} \right)\mathbf{d}x'\mathbf{d}p'\mathbf{d}x} }  - 1,\\
\left\langle {{{\hat x}_A}{{\hat x}_B}} \right\rangle  = \sqrt 2 \cdot {\int {x'x  P\left( {x',p',x} \right)\mathbf{d}x'\mathbf{d}p'\mathbf{d}x} } ,\\
\left\langle {\hat x_B^2} \right\rangle  = \int {{x^2} P\left( {x',p',x} \right)\mathbf{d}x'\mathbf{d}p'\mathbf{d}x},
\end{array}\label{C2}
\end{equation}
where
\begin{equation}
P\left( {x',p',x} \right) = W \cdot {P_{x',p'}} \cdot {\left| {\left\langle {x}
 \mathrel{\left | {\vphantom {x {\sqrt T \alpha }}}
 \right. \kern-\nulldelimiterspace}
 {{\sqrt T \alpha }} \right\rangle } \right|^2},\label{C3}
\end{equation}
and ${P_{x',p'}}$ is ${P_{x_A,p_A}}$ in Eq. (\ref{PS_EQ}) in which $\left\{x_A,p_A\right\}$ are substituted by $\left\{x',p'\right\}$.

After simplifying Eq. (\ref{C2}) by integrating the variable $x$,
\begin{equation}
\begin{array}{l}
\left\langle {\hat x_A^2} \right\rangle  = 2\tilde V - 1,\\
\left\langle {{{\hat x}_A}{{\hat x}_B}} \right\rangle  = 2\sqrt {T} \lambda \tilde V,\\
\left\langle {\hat x_B^2} \right\rangle  = {2T}{\lambda ^2}\tilde V + 1,
\end{array}
\end{equation}
where  $\tilde V = \int {{{x'}^2} \cdot W \cdot {P_{x',p'}}\mathbf{d}x'\mathbf{d}p'}$,
%$\tilde V = \int {{{x'}^2} \cdot Q\left( {\gamma ,\lambda ,T} \right) \cdot {P_{x',p'}}dx'dp'}$,
and further calculation shows
\begin{equation}
\tilde V = \frac{{k + 1}}{{1 - T{\lambda ^2}}}.
\end{equation}

\newpage %Just because of unusual number of tables stacked at end

%\bibliography{reference}% Produces the bibliography via BibTeX.
%\bibliography{QKD-2}% Produces the bibliography via BibTeX.

\begin{thebibliography}{99}

\bibitem{RevModPhys.74.145.2002}
N. Gisin, G. Ribordy, W. Tittel, and H. Zbinden, Rev. Mod. Phys. \textbf{74}, 145 (2002).

\bibitem{RevModPhys.81.1301.2009}
V. Scarani, H. Bechmann-Pasquinucci, N. J. Cerf, M. Du\^{s}ek, N. L\"{u}tkenhaus, and M. Peev, Rev. Mod. Phys.
\textbf{81}, 1301 (2009).

\bibitem{Rev.Mod.Phys.77.513.2005}
S. L. Braunstein and P. van Loock, Rev. Mod. Phys. \textbf{77}, 513 (2005).

\bibitem{Phys.Rep.448.1.2007}
X.-B. Wang, T. Hiroshima, A. Tomita, and M. Hayashi, Phys. Rep. \textbf{448}, 1 (2007).

\bibitem{Rev.Mod.Phys.84.621.2012}
C. Weedbrook, S. Pirandola, R. Garc\'{i}a-Patr\'{o}n, N. J. Cerf, T. C. Ralph, J. H. Shapiro, and S. Lloyd, Rev. Mod.
Phys. \textbf{84}, 621 (2012).

\bibitem{Phys.Rev.Lett.88.057902.2002}
F. Grosshans and P. Grangier, Phys. Rev. Lett. \textbf{88}, 057902 (2002).

\bibitem{Nature.421.238.2003}
F. Grosshans, G. V. Assche, J. Wenger, R. Brouri, N. J. Cerf, and P. Grangier, Nature \textbf{421}, 238 (2003).

\bibitem{Phys.Rev.Lett.93.170504.2004}
C. Weedbrook, A. M. Lance, W. P. Bowen, T. Symul, T. C. Ralph, and P. K. Lam, Phys. Rev. Lett. \textbf{93}, 170504 (2004).

\bibitem{Phys.Rev.Lett.97.190502.2006}
M. Navascu\'{e}s, F. Grosshans, and A. Ac\'{i}n, Phys. Rev. Lett. \textbf{97}, 190502 (2006).

\bibitem{Phys.Rev.Lett.97.190503.2006}
R. Garc\'?a-Patr\'on and N. J. Cerf, Phys. Rev. Lett. \textbf{97}, 190503 (2006).

\bibitem{Phys.Rev.Lett.102.110504.2009}
R. Renner and J. I. Cirac, Phys. Rev. Lett. \textbf{102}, 110504 (2009).

\bibitem{Phys.Rev.Lett.110.030502.2013}
A. Leverrier, R. Garc\'?a-Patr\'on, R. Renner, and N. J. Cerf, Phys. Rev. Lett. \textbf{110}, 030502 (2013).

\bibitem{Phys.Rev.Lett.114.070501.2015}
A. Leverrier, Phys. Rev. Lett. \textbf{114}, 070501 (2015).

\bibitem{Phys.Rev.A.76.042305.2007}
J. Lodewyck, M. Bloch, R. Garc\'ia-Patr\'on, S. Fossier, E. Karpov, E. Diamanti, T. Debuisschert, N. J. Cerf, R. Tualle-Brouri, S. W. McLaughlin, et al., Phys. Rev. A \textbf{76}, 042305 (2007).

\bibitem{New.J.Phys.11.045023.2009}
S. Fossier, E. Diamanti, T. Debuisschert, A. Villing, R. Tualle-Brouri, and P. Grangier, New J. Phys. \textbf{11}, 045023 (2009).

\bibitem{Opt.Express.20.14030.2012}
P. Jouguet, S. Kunz-Jacques, T. Debuisschert, S. Fossier, E. Diamanti, R. All\'eaume, R. Tualle-Brouri, P. Grangier, A. Leverrier, P. Pache, et al., Opt. Express \textbf{20}, 14030 (2012).

\bibitem{Nat.Photonics.7.378.2013}
P. Jouguet, S. Kunz-Jacques, A. Leverrier, P. Grangier, and E. Diamanti, Nat. Photon. \textbf{7}, 378 (2013).

\bibitem{Phys.Rev.A.77.042325.2008}
A. Leverrier, R. All¨¦aume, J. Boutros, G. Z¨¦mor, and P. Grangier, Phys. Rev. A \textbf{77}, 042325 (2008).

\bibitem{Phys.Rev.A.84.062317.2011}
P. Jouguet, S. Kunz-Jacques, and A. Leverrier, Phys. Rev. A \textbf{84}, 062317 (2011).

\bibitem{T.C.Ralph_NLA_QCMC_2009}
T. C. Ralph and A. P. Lund, in Quantum Communication Measurement and Computing Proceedings of 9th
International Conference (AIP, New York 2009), pp. 155 ¨C 160.

\bibitem{Phys.Rev.A.86.012327.2012}
R. Blandino, A. Leverrier, M. Barbieri, J. Etesse, P. Grangier, and R. Tualle-Brouri, Phys. Rev. A \textbf{86}, 012327 (2012).

\bibitem{Phys.Rev.A.87.062311.2013}
B. Xu, C. Tang, H. Chen, W. Zhang, and F. Zhu, Phys. Rev. A \textbf{87}, 062311 (2013).

\bibitem{Nat.Photonics.4.316.2010}
G. Y. Xiang, T. C. Ralph, A. P. Lund, N. Walk, and G. J. Pryde, Nat. Photon. \textbf{4}, 316 (2010).

\bibitem{Phys.Rev.Lett.104.123603.2010}
F. Ferreyrol, M. Barbieri, R. Blandino, S. Fossier, R. Tualle-Brouri, and P. Grangier, Phys. Rev. Lett. \textbf{104}, 123603 (2010).

\bibitem{Phys.Rev.A.83.063801.2011}
F. Ferreyrol, R. Blandino, M. Barbieri, R. Tualle-Brouri, and P. Grangier, Phys. Rev. A \textbf{83}, 063801 (2011).

\bibitem{Nat.Photonics.5.52.2011}
A. Zavatta, J. Fiur¨¢¡¦sek, and M. Bellini, Nat. Photon. \textbf{5}, 52 (2011).

\bibitem{Phys.Rev.A.86.060302.2012}
J. Fiur\'¨¢\^sek and N. J. Cerf, Phys. Rev. A \textbf{86}, 060302 (2012).

\bibitem{Phys.Rev.A.87.020303.2013}
N. Walk, T. C. Ralph, T. Symul, and P. K. Lam, Phys. Rev. A \textbf{87}, 020303 (2013).

\bibitem{Nat.Photonics.8.333.2014}
H. M. Chrzanowski, N. Walk, S. M. Assad, J. Janousek, S. Hosseini, T. C. Ralph, T. Symul, and P. K. Lam, Nat. Photon. \textbf{8}, 333 (2014).

\bibitem{Phys.Rev.A.61.032302.2000}
T. Opatrn\'y, G. Kurizki, and D.-G. Welsch, Phys. Rev. A \textbf{61}, 032302 (2000).

\bibitem{Phys.Rev.A.71.043805.2005}
M. S. Kim, E. Park, P. L. Knight, and H. Jeong, Phys. Rev. A \textbf{71}, 043805 (2005).

\bibitem{Phys.Rev.A.73.042310.2006}
A. Kitagawa, M. Takeoka, M. Sasaki, and A. Chefles, Phys. Rev. A \textbf{73}, 042310 (2006).

\bibitem{Phys.Rev.A.86.012328.2012}
C. Navarrete-Benlloch, R. Garc\'?a-Patr\'on, J. H. Shapiro, and N. J. Cerf, Phys. Rev. A \textbf{86}, 012328 (2012).

\bibitem{Phys.Rev.A.87.012317.2013}
P. Huang, G. He, J. Fang, and G. Zeng, Phys. Rev. A \textbf{87}, 012317 (2013).

\bibitem{Phys.Rev.A.82.062316.2010}
S. L. Zhang and P. van Loock, Phys. Rev. A \textbf{82}, 062316 (2010).

\bibitem{Rev.Sci.Instrum.82.071101.2011}
M. D. Eisaman, J. Fan, A. Migdall, and S. V. Polyakov, Rev. Sci. Instrum. \textbf{82}, 071101 (2011).

\bibitem{Phys.Rev.Lett.96.080502.2006}
M. M. Wolf, G. Giedke, and J. I. Cirac, Phys. Rev. Lett. \textbf{96}, 080502 (2006).

\bibitem{QCQC.Nielsen.2000}
M. A. Nielsen and I. L. Chuang, Quantum Computation and Quantum Communication (Cambridge University Press, Cambridge, 2000).

\bibitem{Phys.Rev.A.83.042312.2011}
A. Leverrier and P. Grangier, Phys. Rev. A \textbf{83}, 042312 (2011).

\bibitem{Phys.Rev.A.81.062343.2010}
A. Leverrier, F. Grosshans, and P. Grangier, Phys. Rev. A \textbf{81}, 062343 (2010).


\end{thebibliography}

\end{document}